\documentclass[usenatbib]{mnras}
\usepackage[utf8]{inputenc}
\usepackage{aas_macros}
\usepackage{amsfonts}
\usepackage{amsmath}
\usepackage{bm}
\usepackage{amssymb}
\usepackage{graphicx}
\usepackage[dvipsnames]{xcolor}
\usepackage{hyperref}
\hypersetup{colorlinks=true,allcolors=teal}
\usepackage{microtype}

\newcommand{\yy}{\ensuremath{\mathbf{y}}}

\renewcommand{\aa}{\ensuremath{\mathbf{a}}}
\renewcommand{\aap}[1]{\ensuremath{\mathbf{a}_{\rm (p)#1}}}
\newcommand{\Fp}[1]{\ensuremath{F_{\rm (p)#1}}}
\newcommand{\ahat}[1]{\ensuremath{\hat{\mathbf{a}}_{#1}}}

\newcommand{\Mpch}{\ensuremath{h^{-1}{\rm Mpc}}}
\newcommand{\Mpc}{\ensuremath{{\rm Mpc}}}

\newcommand{\der}{\ensuremath{{\rm d}}}

\newcommand{\eqn}[1]{equation~\eqref{#1}}
\newcommand{\eqns}[1]{equations~\eqref{#1}}
\newcommand{\ph}[1]{\phantom{#1}}

\newcommand{\be}{\begin{equation}}
\newcommand{\ee}{\end{equation}}
\newcommand{\Cal}[1]{\ensuremath{\mathcal{#1}}}

\title[Bayesian BAO]{Bayesian evidence comparison for distance scale estimates} 
\author[Paranjape \& Sheth]{
 Aseem Paranjape$^{1}$\thanks{E-mail: aseem@iucaa.in} \& 
 Ravi K. Sheth$^{2,3}$\thanks{E-mail: shethrk@physics.upenn.edu},
\\  
 $^1$ Inter-University Centre for Astronomy \& Astrophysics, Ganeshkhind, Post Bag 4, Pune 411007, India\\
 $^2$ Center for Particle Cosmology, University of Pennsylvania, 209 S. 33rd St., Philadelphia, PA 19104, USA\\
 $^3$ The Abdus Salam International Center for Theoretical Physics, Strada Costiera, 11, Trieste 34151, Italy      }

\begin{document}
\label{firstpage}
\pagerange{\pageref{firstpage}--\pageref{lastpage}}
\maketitle

\begin{abstract}
Constraints on cosmological parameters are often distilled from sky surveys by fitting templates to summary statistics of the data that are motivated by a fiducial cosmological model.  However, recent work has shown how to estimate the distance scale using templates that are more generic:  the basis functions used are not explicitly tied to any one cosmological model.  We describe a Bayesian framework for (i) determining how many basis functions to use and (ii) comparing one basis set with another.  Our formulation provides intuition into how 
 (a) one's degree of belief in different basis sets, 
 (b) the fact that the choice of priors depends on basis set, and 
 (c) the data set itself, 
together determine the derived constraints.  
We illustrate our framework using measurements in simulated datasets before applying it to real data.
\end{abstract}


\section{Introduction}
\label{sec:intro}
\noindent 
Using the baryon acoustic oscillation feature in the distribution of galaxies to determine how the cosmological distance scale varies with redshift is one of the primary science drivers of a number of recent or planned galaxy surveys.  There are two approaches:  the first fits a cosmological model-motivated template to measurements of two-point (Fourier or configuration space) correlations.  \cite{cuesta+16} describe such an analysis of the pair correlation function $\xi$ measured in the Baryon Oscillation Spectroscopic Survey \citep[BOSS,][]{BOSSDR12-FinalData}.  However, precise distance scale estimates can also be made without first assuming a fiducial cosmological model \citep{LP2016}.  In \cite{LPboss}, a fifth order polynomial was fit to the same BOSS pair correlation function; this yielded comparable constraints on the distance scale.  

In the second, fiducial cosmology-free approach, there is no compelling reason to have used simple polynomials.  For example, the orthogonal polynomials defined by the eigenvectors of the covariance matrix of $\xi$ are a natural choice.  Recent work has shown that generalised Laguerre functions are also well-motivated in the context of `reconstructing' the shape of $\xi$ \citep{nsz21a}.  This raises the question of how many basis functions are necessary to provide unbiased cosmological constraints.  It is natural to expect the answer to depend on the data -- as data sets improve, higher order polynomials or Laguerre functions will likely be needed \cite[e.g.][]{LPmocks}.  The main goal of our study is to provide a Bayesian evidence-based argument for determining the complexity of the model (e.g., order of polynomial, number of Laguerre functions) which must be fit, as this greatly simplifies such fiducial cosmology-free analyses.  

Section~\ref{sec:bayes} sets up notation and shows how to cast our problem in the Bayesian framework; 
details for deriving the key expression for the Bayesian evidence are provided in an Appendix.
Section~\ref{sec:simrecon} validates our approach using correlation functions measured in cosmological simulations.  It shows that when Bayesian evidence is used to determine the order of the polynomial which should be fit, the subsequent analysis returns unbiased constraints on the (in this case, known) distance scale.  Section~\ref{sec:boss} shows an application to measurements from the BOSS dataset in which the distance scale is, in principle, unknown.  A final section summarizes our results.

\section{Bayesian evidence using the linear Gaussian approximation}\label{sec:bayes}
In this section, we recapitulate some well-known features of Bayesian analysis using the so-called linear Gaussian approximation. We refer the reader to the review by \citet{trotta08} for a more general discussion.

Consider an $N$-dimensional data vector \yy\ with known covariance matrix $C$. Assume that \yy\ is a realisation of a multi-variate Gaussian with unknown mean $\yy^\ast$ and covariance $C$:
\be
\yy \sim \Cal{N}\left(\yy^\ast,C\right)
\ee
We are given a set of $M$ template functions $\{\Cal{T}_m(x)\}_{m=1}^M$ of a control variable $x$, to model the mean using $y^\ast_i \to \sum_{m=1}^M a_m\Cal{T}_m(x_i)$, with the vector \aa\ of linear coefficients to be determined. In other words, under this hypothesis (which we denote \Cal{H}), we assume a Gaussian likelihood for the data:
\be
\yy|_{\aa,\Cal{H}} \sim \Cal{N}\left(\Cal{M}\,\aa,C\right)\,
\ee
where \Cal{M} is the $N\times M$ `design matrix' with elements
\be
\Cal{M}_{im} = \Cal{T}_m(x_i)\,.
\ee
The model is thus linear in the parameter vector \aa, while the templates $\Cal{T}_{m}(x)$ can be arbitrarily nonlinear functions of $x$.

The Bayesian evidence for the hypothesis \Cal{H}, given the observed data set \yy, is
\be
p(\Cal{H}|\yy) = \frac{p(\yy|\Cal{H})\,p(\Cal{H})}{p(\yy)} \label{eq:evidence}
\ee
where $p(\yy|\Cal{H})$ is the probability density of the data 
under the hypothesis \Cal{H} marginalised over all parameter values $\aa$, while $p(\Cal{H})$ is the \emph{a priori} probability attached to the hypothesis, or our degree of belief in the hypothesis in the absence of data. 
In the following, we will only compare two or more models (or hypotheses) given the \emph{same} data set.  This makes the constant $p(\yy)$ irrelevant for what follows, so we will ignore it.

The main problem, then, is to compute $p(\yy|\Cal{H})$.  
If $p(\aa|\Cal{H})$ is the prior probability density of the parameters \aa, and we assume this to also to be an $M$-variate Gaussian with mean $\aap{}$ and inverse covariance \Fp{},
\be
p(\aa|\Cal{H}) \to \Cal{N}\left(\aap{},\Fp{}^{-1}\right)\,,
\ee
then $p(\yy|\Cal{H})$ is an $N$-variate Gaussian, and Appendix~\ref{app:derivn} shows that
\begin{align}
-2\ln p(\yy|\Cal{H}) &= \ln{\rm det}\,C + \yy^{\rm T}C^{-1}\yy + N\ln(2\pi) \notag\\ 
&\ph{\ln}
+ \ln{\rm det}\,\left(F\Fp{}^{-1}\right) - \ahat{}^{\rm T}F\ahat{} + \aap{}^{\rm T}\Fp{}\aap{} \,.
\label{eq:-2lnp(y|H):simple}
\end{align}
Notice that the first line is independent of the model while the second line involves a competition between the \emph{a priori} and \emph{a posteriori} parameter probability densities, where \ahat{} and $F$ are, respectively, the mean and inverse covariance of the latter (see equation~\ref{eq:p(a|y,H)}).
Equation~\eqref{eq:-2lnp(y|H):simple} is sometimes written as 
\begin{align}
-2\ln p(\yy|\Cal{H}) &= \ln{\rm det}\,C + N\ln(2\pi) + \mathbf{e}_y^{\rm T}C^{-1}\mathbf{e}_y \notag\\
&\ph{\ln}
+ \ln{\rm det}\,\left(F\Fp{}^{-1}\right) + \mathbf{e}_p^{\rm T}\Fp{}\mathbf{e}_p \,,
\label{eq:-2lnp(y|H):trad}
\end{align}
where $\mathbf{e}_y\equiv\yy-\Cal{M}\ahat{}$ and $\mathbf{e}_p\equiv\ahat{}-\aap{}$. The first line of \eqn{eq:-2lnp(y|H):trad} now represents the `accuracy' of the model in explaining the data, while the second line, being related to the Kullback-Liebler divergence between the \emph{a priori} and \emph{a posteriori} densities, represents its `complexity'. A `good' model is one in which both these terms are small. Below, however, we will find it more convenient to use and interpret \eqn{eq:-2lnp(y|H):simple}.

In particular, consider two competing hypotheses $\Cal{H}_1$ and $\Cal{H}_2$ for describing the same data \yy. Using \eqn{eq:-2lnp(y|H):simple} in \eqn{eq:evidence} gives
\begin{align}
\ln\left(\frac{p(\Cal{H}_1|\yy)}{p(\Cal{H}_2|\yy)}\right) &=  \frac12\Big[ \ln\left(\frac{{\rm det}\,\Fp{1}}{{\rm det}\,\Fp{2}}\right) \notag\\
 &\ph{+\frac12[]} \quad - \aap{1}^{\rm T}\Fp{1}\aap{1} + \aap{2}^{\rm T}\Fp{2}\aap{2} \Big] \notag\\ 
&\ 
+ \frac12\Big[-\ln\left(\frac{{\rm det}\,F_{1}}{{\rm det}\,F_{2}}\right) + \ahat{1}^{\rm T}F_{1}\ahat{1} - \ahat{2}^{\rm T}F_{2}\ahat{2} \Big] \notag\\
&\ph{+\frac12[]} \quad + \ln\left(\frac{p(\Cal{H}_1)}{p(\Cal{H}_2)}\right)\,.
\label{eq:lnp(H|y):compare}
\end{align}
The above expression lends itself to the following intuitive interpretation.
\begin{itemize}
\item \emph{Degree of belief:}
The last term involving $p(\Cal{H})$ represents a simple relative \emph{a priori} degree of belief for the two hypotheses. To avoid clutter, we will always assume that the models we compare are equally likely (in the absence of data) and therefore set this term to zero. 
\item \emph{Prior comparison:}
The terms in square brackets on the first line represent a comparison of prior assumptions and model construction. 
\begin{itemize}
\item Since each \Fp{}\ is the inverse covariance matrix of the corresponding model, it is clear from the determinant ratio term that small values of ${\rm det}\,\Fp{}$ are disfavoured. Since these could arise due to, both, broad ranges on individual parameters as well as a large number of parameters, this term embodies the concept of Occam's razor by favouring `simpler' models.
\item The terms involving the quadratic form $\aap{}^{\rm T}\Fp{}\aap{}$ disfavour models containing a significant \emph{a priori} departure of the mean parameter values from zero. Thus, these terms represent a penalty for `data-free inference'.
\end{itemize}
\item \emph{Influence of data:}
Finally, the terms in square brackets on the second line (which are identically structured to the first line, but involving the \emph{a posteriori} distribution and containing a crucial relative minus sign) favour significant `detections' of parameter values (i.e., large values of $\ahat{}^{\rm T}F\ahat{}$) accompanied by larger widths of the corresponding covariance matrices (i.e., smaller values of ${\rm det}\,F$). The latter aspect, in particular, is a manifestation of a `goodness-of-fit' criterion, since highly significant detections with low parameter-space volume generally indicate bad fits (e.g., consider fitting a constant to high-quality data drawn from a parabola; the value of the constant will be constrained with small error but will lead to a terrible fit, indicating that more variation in the data must be modelled).
\end{itemize}

\noindent
Motivated by the above, we define the `reduced log-evidence' $\ell({\Cal{H}}|\yy)$ of a hypothesis \Cal{H} given data \yy, as 
\be
\ell({\Cal{H}}|\yy) \equiv \frac12\left[\ln\left(\frac{{\rm det}\,\Fp{}}{{\rm det}\,F}\right) - \aap{}^{\rm T}\Fp{}\aap{} + \ahat{}^{\rm T}F\ahat{}\right]\,,
\label{eq:redlnev}
\ee
so that 
\be
\ln\left(\frac{p(\Cal{H}_1|\yy)}{p(\Cal{H}_2|\yy)}\right) = \ell(\Cal{H}_1|\yy) - \ell(\Cal{H}_2|\yy)
\ee
for models with equal \emph{a priori} degrees of belief $p(\Cal{H}_1)=p(\Cal{H}_2)$. Notice that $\ell(\Cal{H}|\yy)$ does not require explicit calculations of the determinant of the data covariance matrix, making its evaluation numerically more stable than that of \eqn{eq:-2lnp(y|H):trad}.

\subsection{Other estimates}
It is worth mentioning that many analyses in the literature, particularly those that deal with non-Gaussian likelihoods and/or non-linear models, often rely on more approximate statistics such as the (corrected) Akaike Information Criterion (AIC$_c$, below) for model selection. In our language, the AIC$_c$ (\citealp{1974ITAC...19..716A,sugiura}; see \citealp{2007MNRAS.377L..74L} or \citealp{trotta08} for reviews) is given by 
\be
 {\rm AIC}_c =\mathbf{e}_y^{\rm T}C^{-1}\mathbf{e}_y + 2MN/(N-M-1)
\label{eq:AICc}
\ee
(c.f. equation~\ref{eq:-2lnp(y|H):trad}). Similarly to the Bayesian evidence, the AIC$_c$ penalises models with large numbers of parameters, thereby implementing Occam's razor. Although our setup allows us to use the rigorous Bayesian evidence for model comparison, below we will also mention the results of using the AIC$_c$ for comparing the same models.

Finally, in the context of selecting the `simplest, best' model, a useful quantity is the so-called Bayesian complexity $C_b$, given in our setup by \citep[e.g., section 4 of][]{trotta08}
\begin{align}
C_b &\equiv \int\der\aa{}\,p(\aa{}|\yy,\Cal{H})\,
\chi^2(\aa;\yy,\Cal{H})
- \mathbf{e}_y^{\rm T}C^{-1}\mathbf{e}_y \notag\\
&= M - {\rm Tr}\left(F^{-1}\Fp{}\right)\,,
\label{eq:C_b}
\end{align}
where $\chi^2(\aa;\yy,\Cal{H}) \equiv -2\ln[p(\yy|\aa{},\Cal{H})] -\ln\,{\rm det}\,C - N\ln(2\pi)$.
The term ${\rm Tr}\left(F^{-1}\Fp{}\right)$ essentially counts the number of `unconstrained' parameters (for which the width of the \emph{a posteriori} distribution approaches the width of the corresponding \emph{a priori} distribution), so that $C_b$ measures the effective number of parameters in the model. A combination of $C_b$ and $\ell(\Cal{H}|\yy)$ can be useful in breaking degeneracies between models with unequal $M$ that have similar values for $\ell(\Cal{H}|\yy)$, since the inclusion of $C_b$ can help decide whether or not all the parameters in the more complex model are, in fact, required for a good description of the data.

\section{BAO reconstruction:  Validation using simulated datasets}
\label{sec:simrecon}
We noted in the Introduction that, although both simple polynomials and generalized Laguerre functions have been used in the past to furnish unbiased distance scale estimates, as data sets improve, higher order polynomials or Laguerre functions will likely be needed. We show below that our Bayesian analysis provides a simple way to see why, and, further, that when the Bayesian evidence is used, then the resulting constraints on the distance scale are unbiased.

\subsection{Data from simulations}
\label{subsec:sims}
We use the same correlation functions that were presented in \cite{OTrec}.  
These were obtained by identifying halos with masses greater than $1.3\times 10^{13}h^{-1}M_\odot$ using a friends of friends algorithm with link-length parameter 0.2, in $N_{\rm real,max}=20$ independent realisations of the HADES simulation suite \citep{vn+18}.  The cosmological parameters of the simulation were $\Omega_{\rm m}=0.3175, \Omega_{\rm b}=0.04586, \Omega_{\Lambda}=0.6825, n_{\rm s}=0.9624, h=0.6711, \sigma_8=0.833$.
In each realization, $\xi$ of the halos was measured in bins of width $1\Mpch$ over the range $60\leq s/(\Mpch) \leq 120$.  The linear bias parameter, $b=\sqrt{\xi_{\rm halos}/\xi_{\rm dm,Lin}}$ on these large scales is approximately $1.3$.  

\begin{figure*}
\centering
\includegraphics[width=0.47\textwidth,trim=20 0 10 20 clip]{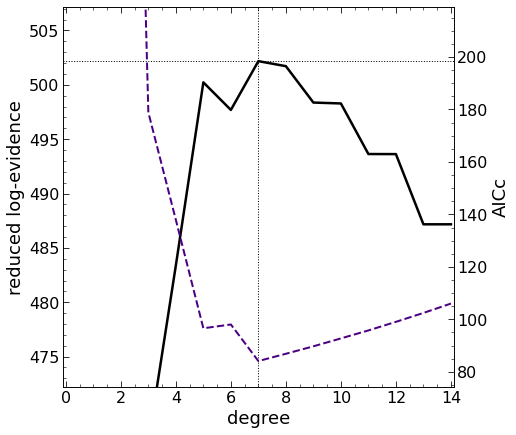}
\includegraphics[width=0.45\textwidth,trim=0 20 0 0 clip]{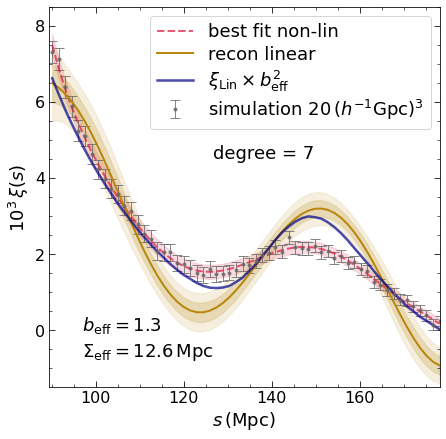}
\caption{Bayesian model selection using halo 2pcf measurements combining $N_{\rm real}=20$ realisations of the HADES simulation. 
\emph{(Left panel:)} Reduced log-evidence $\ell(\Cal{H}|\yy)$ (equation~\ref{eq:redlnev}) as a function of Laguerre function degree (solid black curve marked by the left vertical axis). We select degree 7 (vertical dotted line) as the `best, least complex' model describing the data, with the maximum $\ell(\Cal{H}|\yy)$ indicated by the horizontal dotted line. The dashed purple line, marked by the right vertical axis, shows the more frequentist AIC$_c$ statistic, which also selects degree $7$ in this case. \emph{(Right panel:)} Comparison of model and data. Grey points with errors show the measured non-linear real space 2pcf $\xi(s)$ of haloes having mean linear bias $b=1.3$. We display $10^3\xi(s)$ for the reasons discussed in the text. Red dashed curve shows the degree 7 Laguerre function fit \eqref{eq:xiNL(s)} using the \emph{a posteriori} mean \ahat{} and smearing scale $\Sigma=8.45\Mpch=12.6\,\Mpc$. Red band shows the result of sampling the full \emph{a posteriori} distribution $p(\aa|\yy,\Cal{H})$ and constructing the $16^{\rm th}$ and $84^{\rm th}$ percentiles of the predicted 2pcf at each scale $s$, i.e., the predicted central $68\%$ confidence region in data space. Yellow solid curve shows the reconstructed linear 2pcf $\xi_{\rm L}(r)$ obtained by inserting \ahat{} into \eqn{eq:xiL(r)}. Inner and outer yellow bands respectively show the corresponding predicted central $68\%$ and $95\%$ confidence regions obtained by sampling $p(\aa|\yy,\Cal{H})$. Thick blue solid curve shows the theoretical prediction for $b^2\xi_{\rm Lin}(r)$. See text for a discussion.
}
\label{fig:results}
\end{figure*}

Our primary statistic is the arithmetic mean of $\xi(s)$ in each $s$ bin, over $N_{\rm real}\leq N_{\rm real,max}$ of these realisations.  The grey symbols with error bars in the right hand panel of Fig.~\ref{fig:results} show this mean for the full set ($N_{\rm real}=20$) which gives an equivalent volume of $20\, (h^{-1}{\rm Gpc})^3$.  The symbols with larger error bars in Fig.~\ref{fig:results_2real} show $\xi$ averaged over $N_{\rm real}=2$ randomly chosen realizations whose equivalent volume of $2\, (h^{-1}{\rm Gpc})^3$ is similar to (but somewhat larger than) the effective volume of the CMASS sample of the BOSS survey \citep{cuesta+16}.  In each case, our data set \yy\ comprises the $N=60$ values of this averaged 2pcf.

To proceed, we need the covariance matrix $C$.  In principle, one could estimate it from the different realizations.  In practice, 20 realizations is too few to provide a reliable estimate.  Previous work has shown that $C^{\rm (lin)}$, a linearly biased linear theory + Poisson shot noise model is appropriate for estimating the covariance matrix on scales larger than about $60\Mpch$ \citep{LPnus}.  
If we compute $C^{\rm (lin)}$ for a single realization, then $C$ for $N_{\rm real}$ independent realisations, an effective volume that is $N_{\rm real}$ times larger, is given by $C=C^{\rm (lin)}/N_{\rm real}$.

\subsection{Laguerre reconstruction setup}
\cite{nsz21a} show that if the linear theory 2pcf $\xi_{\rm L}(r)$ around the BAO feature is described by a polynomial of degree $M-1$ (i.e., an $M$-dimensional model) with coefficients $\aa = \{a_m\}_{m=0}^{M-1}$,
\be
 \xi_{\rm L}(r) = \sum_{m=0}^{M-1}\,\frac{a_m}{m!}\left(\frac{r-r_{\rm fid}}{\sigma_{\rm fid}}\right)^m\,,
\label{eq:xiL(r)}
\ee
then the non-linear 2pcf $\xi(s)$ should be very well approximated by the Laguerre expansion
\be
\xi(s) = \sum_{m=0}^{M-1}\,\frac{a_m}{m!}\left(\frac{\Sigma}{\sigma_{\rm fid}}\right)^m \, \nu_m\left(\frac{s}{\Sigma};\frac{r_{\rm fid}}{\Sigma}\right)\,,
\label{eq:xiNL(s)}
\ee
where the functions $\nu_m(x;x_{\rm fid})$ are the `centered' generalized Laguerre functions defined in Appendix~B of \cite{nsz21a}.

Our goal is to determine the best choice of $M$, and the $M$ coefficients $\aa = \{a_m\}_{m=0}^{M-1}$ associated with fitting equation~(\ref{eq:xiNL(s)}) to the measurements.  
However, before performing the fitting exercise, we must make some choices regarding three length scales:  
$r_{\rm fid}$ (a centering separation), $\sigma_{\rm fid}$ (a fiducial normalisation) and $\Sigma$ (a `smearing scale' that is related to the physics of gravitational evolution).  

The values of $r_{\rm fid}$ and $\sigma_{\rm fid}$ do not affect the quality of the fit, but can improve the numerical stability of the fitting exercise and change the correlation structure of the \emph{a posteriori} parameter distribution. As such, they should generally be chosen using the typical length scales seen in the data. Throughout, we will set $r_{\rm fid}=90\Mpch$ (approximately the mid-point of the BAO dip-and-peak feature) and $\sigma_{\rm fid}=10\Mpch$ (approximately half the width of the feature), having checked that reasonable variations around these values have no impact on our results. As regards the third scale $\Sigma$, although its value could be left as a free parameter to be determined from the same data set used for the Laguerre BAO reconstruction \citep{nsz22}, for the analysis of simulations in this work we simply fix $\Sigma$ to the value expected from linear theory $8.45\Mpch=12.6\Mpc$ \cite[see the discussion in][]{nsz21a}. 

Finally, our choice of including factorials explicitly in the model definitions~\eqref{eq:xiL(r)} leads to better numerical stability when exploring high-dimensional models ($M\gtrsim10$). However, to retain the usual intuition of polynomial behaviour, we must be careful in setting priors on the parameter values, as we discuss next.

\subsubsection{Choice of prior mean and covariance}
Standard least-squares fitting of data using polynomial models $\sum_m c_m\,x^m$ typically assumes uninformative priors on the coefficients $c_m$. In our case, since $a_m\sim c_m\,(m!)$, this effect can be mimicked by assuming a prior on $a_m$ which broadens proportionally to $m!$. 
Moreover, as we saw in section~\ref{sec:bayes}, the Gaussianology formalism is analytically tractable (and particularly simple) when using Gaussian priors.

To this end, we first assume \Fp{} to be a diagonal matrix, with diagonal entries $1/(m!)^2$. Next, visual inspection of the data suggests an approximately sinusoidal behaviour of the BAO feature with amplitude $A\sim2\times10^{-3}$, so we assume $a_{{\rm (p)}m} = (-1)^{m/2}A$ for even $m$ and zero for odd $m$ (which defines a Taylor series for $A\,\sin(x)$ truncated at order $M-1$) and correspondingly multiply \Fp{} by $A^{-2}$. To approach the `uninformative' limit, we further multiply \Fp{} by a factor $10^{-2}$, so that the non-zero values of \aap{} represent only $\pm0.1\sigma$ deviations from zero. We have also checked that varying this last factor so as to make these mean values $\pm0.03\sigma$ or $\pm0.3\sigma$ deviations from zero, makes no qualitative difference to our final results. At this stage, then, our \emph{a priori} distribution is characterised by
\begin{align}
\aap{} &= A\times\left(1,0,-1,0,\ldots\right)\,,\notag\\
\Fp{}^{-1} &= 10^2\times A^2\times{\rm diag}\left\{(m!)^2\right\}_{m=0}^{M-1}\,,
\end{align}
with $A=2\times10^{-3}$. 

Additionally, since the $\xi(s)$ values in the range we probe are numerically all $\ll1$, the numerical stability of the fitting exercise improves if we scale the data vector $\yy\to S\times\yy$ by a constant factor $S\simeq10^2$-$10^3$. Since $S$ has no physical or statistical relevance, the final choice of the shape of the fitted function should be explicitly independent of $S$. Inspection of \eqns{eq:F-def} and~\eqref{eq:ahat-def} shows that, since \ahat{} is linear in the data \yy, we must then also scale $\aap{}\to S\times\aap{}$ and $\Fp{}^{-1}\to S^2\times\Fp{}^{-1}$. We will display results for $S=10^3$ below, after performing this rescaling.

Finally, all of the above assumes that the `uninformative' behaviour of \Fp{} is valid at a specified fiducial normalisation $\sigma_{\rm fid}$ (we motivated the value $\sigma_{\rm fid}=10\Mpch$ above). Having made this choice defining the structure of \Fp{}, if we now decide to work at a different fiducial scale $\sigma_{\rm fid}^\prime$, then the corresponding change in the definition of the parameters \aa{} requires a further scaling $a_{{\rm (p)}m}\to (\sigma_{\rm fid}^\prime/\sigma_{\rm fid})^m\times a_{{\rm (p)}m}$ and $(F^{-1}_{\rm (p)})_{mm}\to (\sigma_{\rm fid}^\prime/\sigma_{\rm fid})^{2m}\times (F^{-1}_{\rm (p)})_{mm}$. While we implement this in our code, we note that there is no particular reason to demand the $(m!)^2$ behaviour of $\Fp{}^{-1}$ at $\sigma_{\rm fid}=10\Mpch$. We therefore consider this as the main \emph{ad hoc} assumption of our model, and show all results for $\sigma_{\rm fid}=10\Mpch$. We have checked that small variations in this default scale at which the $(m!)^2$ scaling is defined do not change any of our results. We have also verified that setting all elements of \aap{} to zero, while keeping the other choices intact, makes no difference to our results.

\begin{figure}
\centering
\includegraphics[width=0.45\textwidth]{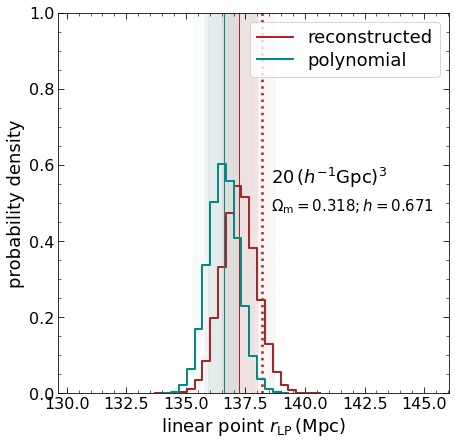}
\caption{Predicted distributions for the BAO linear point $r_{\rm LP}$ after Laguerre recontruction (red histogram) and after fitting a polynomial to the nonlinear measurements (cyan histogram), obtained by sampling the \emph{a posteriori} distribution $p(\aa|\yy,\Cal{H})$ from the fitting exercise described in Fig.~\ref{fig:results} and then applying the procedure described in the text. For each distribution, the vertical solid line shows the median, while the inner and outer bands respectively show the  central $68\%$ and $95\%$ confidence regions, and the vertical dotted red line shows the linear theory prediction. See text for a discussion.}
\label{fig:scales}
\end{figure}

\subsection{Results}
Fig.~\ref{fig:results} shows the results of the Bayesian selection exercise for $N_{\rm real}=20$. The \emph{left panel} shows the reduced log-evidence $\ell(\Cal{H}|\yy)$ for Laguerre function fits of varying degree. We see that degrees 7 and 8 are clearly preferred over all others, with degrees 5, 6 and 9 being rejected with `substantial' strength of evidence as per the Jeffreys scale, with all others rejected `strongly' or `decisively'. The sharp decline at small degrees is almost entirely due to `bad fits', while the decline at high degrees reflects the increasing penalty for more `complex' models (see the discussion below equation~\ref{eq:lnp(H|y):compare}). The Bayesian complexity $C_b=7.94$ for the best-fit degree 7 function is close to the dimension $M=8$ of the model, while $C_b=8.35$ for the best-fit degree 8 function is somewhat smaller than the dimension $M=9$ of the model.  The degree 7 Laguerre function therefore provides a statistically well-motivated description of the data.

The \emph{right panel} of Fig.~\ref{fig:results} compares the data (grey points with error bars) with the model prediction derived from the \emph{a posteriori} distribution $p(\aa{}|\yy,\Cal{H})$ for the degree 8 Laguerre function (red dashed line with red error band; this is the projection of $p(\aa{}|\yy,\Cal{H})$ into data space using $\aa{}\to\Cal{M}\,\aa{}$). Clearly, the model provides an excellent description of the data. We can now exploit the fact that the \emph{same} distribution $p(\aa{}|\yy,\Cal{H})$, when applied to \eqn{eq:xiL(r)} instead of \eqn{eq:xiNL(s)}, should provide a description of the linear theory 2pcf. The yellow solid line with yellow bands shows this prediction of the `reconstructed' linear 2pcf, while the thick solid blue curve shows the actual linear theory result $b^2\xi_{\rm Lin}(r)$. We see reasonably good agreement in the vicinity of the peak, although the reconstruction deviates from the true function at separations $\gtrsim20\,\Mpc$ away from the peak.
This departure is likely due to the various approximations inherent in deriving \eqn{eq:xiNL(s)} from \eqn{eq:xiL(r)}: (i) the use of the linear theory estimate for $\Sigma$, (ii) ignoring mode coupling and (iii) ignoring scale-dependent bias \citep[see the discussion in][]{nsz21a}. Additionally, the presence of cosmic variance makes the shape of the linear 2pcf in individual simulation realisations differ from its mean value  $b^2\xi_{\rm Lin}$ which we have used.

We have also explicitly checked that using the \emph{a posteriori} distributions for the higher degree Laguerre functions does not affect the \emph{right panel} of Fig.~\ref{fig:results}, apart from broadening the yellow bands at separations $\gtrsim150\,\Mpc$, confirming that the Bayesian selection of degree 7 indeed produces the `best, least complex fit' to the data.\footnote{We have also checked that replacing the Laguerre functions with simple polynomials gives nearly identical results for the reduced log-evidence, with only numerical differences between the best-fit parameter values. The use of Laguerre functions can then be ascribed to a theoretical prior $p(\Cal{H}_{\rm Laguerre}) > p(\Cal{H}_{\rm polynomial})$ for describing $\xi(s)$.} 

% HADES linear theory: (r_pk, r_LP) = (148.88,138.19) Mpc

The polynomial approximation to the linear 2pcf in \eqn{eq:xiL(r)} can be used, for any given parameter vector \aa{}, to numerically estimate the scales $r_{\rm peak}$ and $r_{\rm dip}$ corresponding to the maximum and minimum, respectively, of the reconstructed BAO feature (i.e., the roots of $\der\xi_{\rm L}/\der r$). In terms of these roots, the linear point \citep{LP2016} is given by $r_{\rm LP} = (r_{\rm peak}+r_{\rm dip})/2$. The histograms in Fig.~\ref{fig:scales} show the distribution of the scales $r_{\rm LP}$ inferred by applying this procedure to a sample from the \emph{a posteriori} distribution $p(\aa{}|\yy,\Cal{H})$. The red histogram shows the result for the Laguerre reconstruction, while the cyan shows the result of directly fitting $\xi(s)$ with a degree 5 polynomial. The vertical dotted red line indicates the theoretical value $r_{\rm LP}=138.2\,\Mpc$ for the HADES cosmology. The vertical solid lines and inner bands indicate the median and central $68\%$ confidence region of the inferred values, respectively $r_{\rm LP} = 137.2^{+0.7}_{-0.7}\Mpch$ (Laguerre reconstruction) and $r_{\rm LP} = 136.6^{+0.6}_{-0.6}\Mpch$ (polynomial fit). The outer (fainter) vertical bands correspondingly show the central $95\%$ confidence region. 
We see that the theoretical value of $r_{\rm LP}$ 
lies within the central $95\%$ confidence region 
and is $\sim0.7\%$ away from the median of the recovered values using Laguerre reconstruction, while the polynomial fit prefers slightly lower values, as expected from the discussion in \citet{LP2016}.\footnote{In principle, one could also envisage estimating an averaged linear point, calculated by combining the linear point \emph{a posteriori} distributions for each $M$, weighted by the Bayesian evidence. We have chosen not to do this in this work.}

% N_real=20 60-120
% reconstructed
% central 68%
% r_pk : 150.75 + 0.83 - 0.77 Mpc
% r_dip: 123.73 + 0.79 - 0.83 Mpc
% r_LP : 137.23 + 0.73 - 0.71 Mpc
% central 95%
% r_pk : 150.75 + 1.72 - 1.48 Mpc
% r_dip: 123.73 + 1.53 - 1.77 Mpc
% r_LP : 137.23 + 1.46 - 1.42 Mpc

% nonlinear-poly
% central 68%
% r_LP : 136.58 + 0.65 - 0.64 Mpc
% central 95%
% r_LP : 136.58 + 1.30 - 1.27 Mpc

% N_real=20 60-130
% reconstructed
% central 68%
% r_pk : 149.24 + 0.97 - 0.90 Mpc
% r_dip: 122.25 + 0.77 - 0.77 Mpc
% r_LP : 135.74 + 0.80 - 0.75 Mpc
% central 95%
% r_pk : 149.24 + 2.02 - 1.71 Mpc
% r_dip: 122.25 + 1.56 - 1.54 Mpc
% r_LP : 135.74 + 1.65 - 1.46 Mpc

% nonlinear-poly
% central 68%
% r_LP : 136.13 + 0.65 - 0.65 Mpc
% central 95%
% r_LP : 136.13 + 1.31 - 1.27 Mpc

\begin{figure*}
\centering
\includegraphics[width=0.45\textwidth,trim=20 0 10 20 clip]{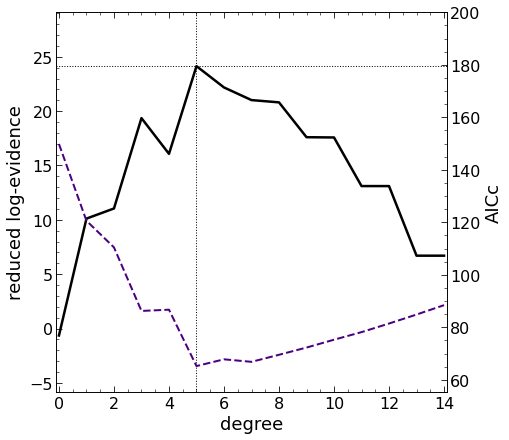}
\includegraphics[width=0.45\textwidth,trim=0 20 0 0 clip]{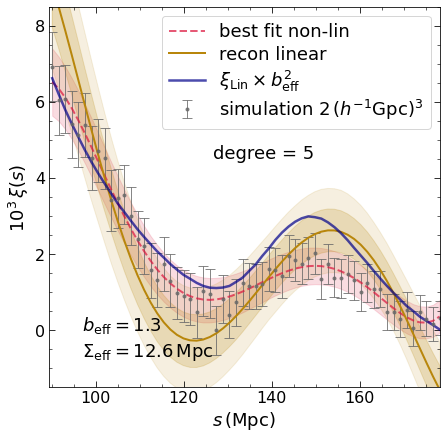}
\caption{Same as Fig.~\ref{fig:results}, using $N_{\rm real}=2$ realisations, corresponding to an effective simulation volume similar to the BOSS CMASS sample. In this case, the Bayesian selection favours a degree 5 Laguerre function, as does the AIC$_c$ statistic.}
\label{fig:results_2real}
\end{figure*}

\begin{figure}
\centering
\includegraphics[width=0.45\textwidth]{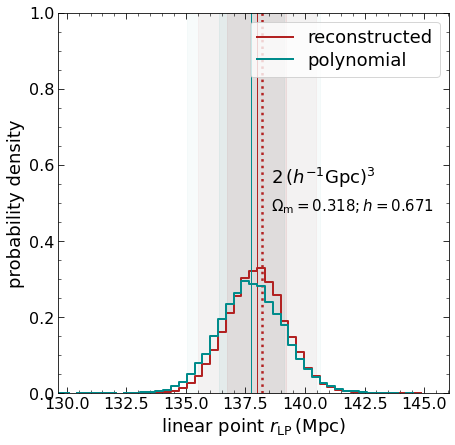}
\caption{Same as Fig.~\ref{fig:scales}, using \emph{a posteriori} parameter distribution selected using $N_{\rm real}=2$ realisations (see Fig.~\ref{fig:results_2real}), corresponding to an effective simulation volume similar to the BOSS CMASS sample.}
\label{fig:scales_2real}
\end{figure}

It is also very interesting to note that, had we forced a reconstruction with a higher or lower degree Laguerre function than the degree 7 selected by our Bayesian analysis, the recovery of $r_{\rm LP}$ would be substantially degraded. For example, using the `less complex' degree 5 function leads to a very similar width of the $r_{\rm LP}$ distribution, but centered on $r_{\rm LP}=136.2\,\Mpc$, thus excluding the true value at $>95\%$ confidence. On the other hand, using the `more complex' degree 9 Laguerre function gives exactly the same median $r_{\rm LP}$ value, but with a width that is nearly twice as large. 
Similar behaviour is seen with smaller volume samples, with the Bayesian selection now favouring lower degree Laguerre functions. Thus, the Bayesian selection approach has the very practical benefit of optimising the estimate on $r_{\rm LP}$, in terms of minimising its bias as well as error.

Figs.~\ref{fig:results_2real} and~\ref{fig:scales_2real} are formatted identically to Figs.~\ref{fig:results} and~\ref{fig:scales}, respectively, and show the results when repeating the exercise with $N_{\rm real}=2$ realisations, so as to approximately mimic the volume probed by the BOSS CMASS sample. This time, the degree 5 Laguerre function is preferred by the log-evidence comparison, which is sensible, given the noisier data. 
The selection of degree 5 is also relatively clean already with $\ell(\Cal{H}|\yy)$ (unlike the case in Fig.~\ref{fig:results}), so that the Bayesian complexity is not additionally required. Nevertheless, we have checked that $C_b=5.995$ for the best-fit model in this case, consistent with the dimension $M=6$.  The prediction using the corresponding best fit parameters describes the data very well, while the quality of the linear reconstruction is now relatively degraded, with larger deviations of the median prediction from the true linear theory behaviour (which is nevertheless within the, now broader, $95\%$ confidence interval predicted by the \emph{a posteriori} distribution). 

We also see that $r_{\rm LP}$ is recovered very well, with a median and central $68\%$ confidence interval of $r_{\rm LP} = 138.0^{+1.2}_{-1.2}\,\Mpc$ using Laguerre reconstruction, with the median being very close to the true value, while the same using the polynomial fit to $\xi(s)$ is $r_{\rm LP} = 137.8^{+1.4}_{-1.4}\,\Mpc$, i.e. favouring slightly smaller values as expected, but only marginally in this case.

% N_real=2, pair 1,2 60-120
% reconstructed
% central 68%
% r_pk : 153.42 + 1.48 - 1.48 Mpc
% r_dip: 122.50 + 1.53 - 1.49 Mpc
% r_LP : 137.97 + 1.22 - 1.23 Mpc
% central 95%
% r_pk : 153.42 + 3.01 - 3.10 Mpc
% r_dip: 122.50 + 3.21 - 2.98 Mpc
% r_LP : 137.97 + 2.51 - 2.46 Mpc

% nonlinear-poly
% central 68%
% r_LP : 137.75 + 1.38 - 1.36 Mpc
% central 95%
% r_LP : 137.75 + 2.88 - 2.71 Mpc

% N_real=2, pair 1,2 60-130
% reconstructed
% central 68%
% r_pk : 150.23 + 2.26 - 2.07 Mpc
% r_dip: 119.93 + 1.84 - 1.57 Mpc
% r_LP : 135.09 + 1.78 - 1.55 Mpc
% central 95%
% r_pk : 150.23 + 4.93 - 4.06 Mpc
% r_dip: 119.93 + 4.20 - 3.00 Mpc
% r_LP : 135.09 + 3.94 - 2.95 Mpc

% nonlinear-poly
% central 68%
% r_LP : 136.58 + 1.90 - 1.65 Mpc
% central 95%
% r_LP : 136.58 + 4.06 - 3.21 Mpc

\begin{figure*}
\centering
\includegraphics[width=0.45\textwidth,trim=20 0 10 20 clip]{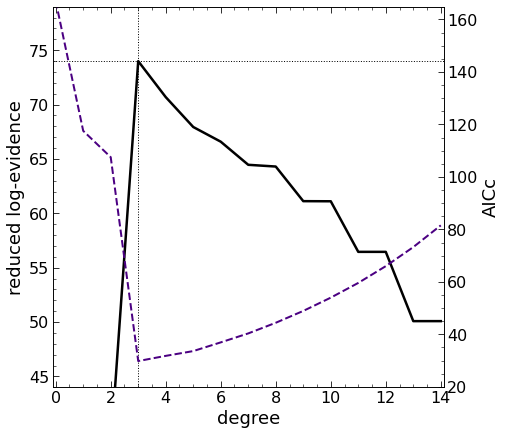}
\includegraphics[width=0.45\textwidth,trim=0 20 0 0 clip]{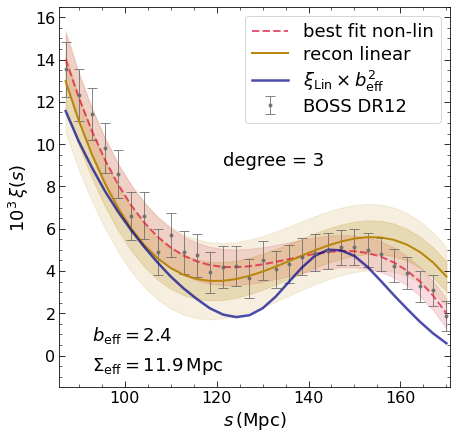}
\caption{Same as Fig.~\ref{fig:results}, showing results for the BOSS CMASS sample. The Bayesian selection now favours a degree 3 Laguerre function, as does the AIC$_c$ statistic.}
\label{fig:results_boss}
\end{figure*}

\section{Bayesian BAO reconstruction in BOSS}
\label{sec:boss}
We have repeated this analysis using the redshift-space `pre-reconstruction' monopole $\xi_0(s)$ measured by \citet{cuesta+16} in the CMASS sample (median redshift $z=0.57$) of the BOSS DR12 data \citep{BOSSDR12-FinalData}.\footnote{Strictly speaking, the publicly available measurements are reported in bins of width $4h^{-1}$Mpc, whereas the analysis below uses measurements in narrower bins (of width $2h^{-1}$Mpc) that were kindly provided by A. Cuesta.  We will, nevertheless, refer to them as the measurements of \citet{cuesta+16}.
Also, while we do not pursue this, the formalism can just as easily be applied to fitting simple polynomials to `post-reconstruction' estimates of the 2pcf.
}
Our goal here is to provide a proof-of-concept of our technique on actual data. We therefore fix all the theoretical cosmology-dependent factors in our setup (such as the smearing scale $\Sigma$ and the expected linear theory value of $r_{\rm LP}$) to the values expected in the flat $\Lambda$CDM cosmology used as the fiducial model in the BOSS DR12 final cosmological analysis \citep{BOSSDR12-FinalCosmo}: $\Omega_{\rm m}=0.31, \Omega_{\rm b}=0.04814, \Omega_{\Lambda}=0.69, n_{\rm s}=0.97, h=0.676, \sigma_8=0.8$. We will refer to this as the BOSS cosmology below. The values of $\Omega_{\rm m}$ and $h$ in the fiducial cosmology are nearly identical to the corresponding best-fit values derived from the final constraints using BOSS DR12 by \citet{BOSSDR12-FinalCosmo}. 

This cosmology is slightly different from the fiducial cosmology assumed by \citet{cuesta+16} to convert observed angles and redshifts into distances. This flat $\Lambda$CDM cosmology, having $\Omega_{\rm m}=0.29, \Omega_{\rm b}=0.04586, \Omega_{\Lambda}=0.71, n_{\rm s}=0.97, h=0.7, \sigma_8=0.8$, was the same as in the QPM mock halo catalogs \citep{wtm14} used by \citet{cuesta+16} for calculating the covariance matrix of the $\xi_0$ measurements. We will refer to this as the QPM cosmology below. In order to be consistent, we will report all derived numbers after converting length scales from \citet{cuesta+16} into equivalent ones in the BOSS cosmology using the rescaling $\ell_{\rm BOSS} = \ell_{\rm QPM}\times D_{V,{\rm BOSS}}/D_{V,{\rm QPM}}$ for any length scale $\ell$, where $D_V$ is the volume averaged distance scale given by equation~(6) of \citet{cuesta+16}, and also self-consistently use Mpc units throughout.

The monopole measurements are provided in $2/0.7\,\Mpc$ wide bins. The range of data chosen for the fitting exercise is known to have small effects on the final result \cite[e.g.][]{LPmocks}. One can then imagine a Bayesian evidence exercise in which one marginalises over a prior on the choice of this range. In the present work, since our focus is on presenting a proof-of-concept analysis, we simply restrict to the range $60/0.7\leq s/(\Mpc)\leq120/0.7$ \citep[using the measurements provided by][]{cuesta+16}, giving us $N=30$ data points, and comment on the sensitivity of our results to this choice in Appendix~\ref{app:scales}. We will report the results of a full Bayesian analysis marginalised over the range of scales in a separate work.

For the covariance matrix, we use mock measurements averaged over 1000 realisations of the QPM mock halo catalogs mentioned above. In principle, the covariance matrix could be slightly different in the BOSS cosmology that we have adopted as our theory reference. For simplicity, we ignore this difference.

The haloes selected for the QPM mocks had a real space linear bias of $b=2.1$. Since the value of $\sigma_8$, which is the main cosmological parameter degenerate with the value of $b$, is the same in the BOSS and QPM cosmologies, we can simply adopt $b=2.1$ as the value of real space halo bias.
Since we are working in redshift space, the calculation of halo bias $b$ and the smearing scale $\Sigma(z)$ must also account for redshift space distortions at large scales. We follow \citet{nsz22} and replace $b\to b_{\rm eff}$ and $\Sigma\to\Sigma_{\rm eff}$, by first defining $\beta=f/b$ (where $f=\der\ln D/\der\ln a = 0.784$ and $D(z)/D(0)=0.744$ at $z=0.57$ for the BOSS cosmology) and then writing
\begin{align}
b_{\rm eff}^2 &= b^2\left(1+\frac{2\beta}3 + \frac{\beta^2}5\right) \simeq (2.4)^2\,,\\
\Sigma_{\rm eff}^2 &= \Sigma^2\left[1+\frac{f}3\left(2+f\right) \frac{\left(1+6\beta/5+3\beta^2/7\right)}{\left(1+2\beta/3+\beta^2/5\right)}\right] \notag\\
&\simeq \left(11.9\,\Mpc\right)^2\,.
\end{align}
We use this setup and proceed in the same way as in section~\ref{sec:simrecon}.
Fig.~\ref{fig:results_boss} is formatted identically to Fig.~\ref{fig:results} and shows the results for the CMASS measurements. We see a clear preference for a degree 3 Laguerre function, which gives an excellent description of the $\xi_0(s)$ data. 
As with the small-volume simulation results of Fig.~\ref{fig:results_2real}, the value of $C_b=3.998$ for the best-fit is consistent with the dimension $M=4$, but is not needed since $\ell(\Cal{H}|\yy)$ has a well-defined peak at degree 3.
The BAO peak is not very pronounced, however, and the corresponding linear reconstruction overestimates the expected peak location in the BOSS cosmology. 
In this case, in addition to the caveats mentioned above in the context of the real-space fits, the value of halo bias $b$ in the data is much more uncertain (e.g., due to stochasticity of the galaxy-halo connection). Our intention in showing the blue curve is therefore simply to guide the eye. We could alternatively have chosen $b_{\rm eff}$ such that $b_{\rm eff}^2 \xi_{\rm Lin}$ has the same amplitude as $\xi_{\rm L}$ either around the linear point, or at $\sim 100 {\rm Mpc}/h$, which would improve the visual agreement but not affect any of our conclusions.

\begin{figure}
\centering
\includegraphics[width=0.45\textwidth]{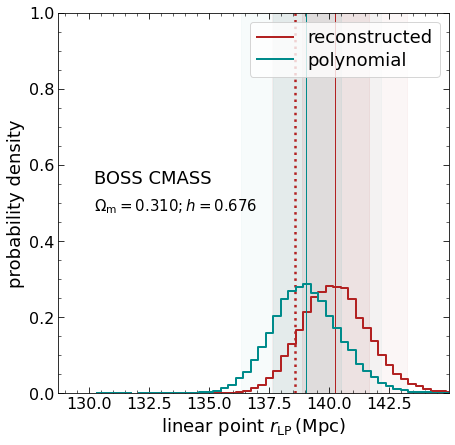}
\caption{Same as Fig.~\ref{fig:scales}, using the \emph{a posteriori} parameter distribution selected using the BOSS CMASS sample (see Fig.~\ref{fig:results_boss}).}
\label{fig:scales_boss}
\end{figure}

Fig.~\ref{fig:scales_boss} (which is formatted identically to Fig.~\ref{fig:scales}) shows that, as noted in earlier works, the recovery of the linear point $r_{\rm LP}$ is relatively much more stable. We find a median and central $68\%$ confidence interval of $r_{\rm LP} = 140.2^{+1.4}_{-1.3}\,\Mpc$ using the Laguerre reconstruction (the corresponding value from the polynomial fit to $\xi$ is $r_{\rm LP} = 139.0^{+1.5}_{-1.4}\,\Mpc$). The expected theoretical value of $r_{\rm LP}=138.6\,\Mpc$ in the BOSS cosmology and is well within the central $95\%$ confidence interval of the Laguerre reconstruction.\footnote{The median value recovered from fitting a polynomial to $\xi$ is slightly different from the value reported by \citet{LPboss} using the same measurements and covariance matrix. Appendix~\ref{app:scales} shows that this small (statistically insignificant) discrepancy arises from the small difference in the range of scales used by those authors.} 

% BOSS  linear theory: (r_pk, r_LP) = (149.37,138.55) Mpc

% BOSS 60-120
% reconstructed
% central 68%
% r_pk : 157.66 + 1.78 - 1.97 Mpc
% r_dip: 122.77 + 2.56 - 2.19 Mpc
% r_LP : 140.22 + 1.44 - 1.34 Mpc
% central 95%
% r_pk : 157.66 + 3.53 - 4.28 Mpc
% r_dip: 122.77 + 5.67 - 4.11 Mpc
% r_LP : 140.22 + 3.00 - 2.59 Mpc

% nonlinear-poly
% central 68%
% r_LP : 139.01 + 1.48 - 1.35 Mpc
% central 95%
% r_LP : 139.01 + 3.12 - 2.68 Mpc

\section{Discussion}\label{sec:discuss}
We have presented a Bayesian approach to performing Laguerre reconstruction \citep{nsz21a} and inferring the cosmological distance scale $r_{\rm LP}$ \citep{LP2016} from measurements of the 2pcf. Our framework exploits the linearity of model parameters in the Laguerre reconstruction technique, leading to a fully analytical calculation of the Bayesian evidence. Standard ideas from Bayesian statistics then allow for a principled selection of the degree of the Laguerre function, as well as the \emph{a posteriori} probability distribution of its parameters, that provides the `best, least complex' description of 2pcf data.

We applied these ideas to, both, simulated dark matter haloes as well as the redshift-space galaxy 2pcf monopole $\xi_0(s)$ derived from BOSS DR12 data in rectangular bins of separation $s$. Given the number of approximations made in our analysis (such as neglecting mode coupling and scale dependent bias when calculating the smearing scale $\Sigma$ used in equation~\ref{eq:xiNL(s)}), the fact that the Laguerre-reconstructed value of $r_{\rm LP}$ from the BOSS CMASS sample agrees with the theoretical expectation in the BOSS final cosmology to $\sim1\%$ is very encouraging, and motivates an extension of our technique to include the full anisotropy of the redshift space 2pcf. This will be particularly interesting for upcoming larger volume samples such as DESI \citep{DESI}. For such samples, it will also be important to extend our framework to include a fully Bayesian treatment of the range of $s$ values that should be used in the fitting analysis; this is work in progress. 

Finally, although our analysis fit (non-central) generalised Laguerre functions to $\xi(s)$ data provided in rectangular bins of $s$, it applies equally to the case in which $\xi$ is directly measured in bins shaped according to the Laguerre polynomials (Al Rahman et al, in preparation). 
In addition, although we have focused on modelling $\xi$ on large BAO-relevant scales, recent work has highlighted the usefulness of a polynomial parametrization on small nonlinear scales \citep{smallr}.  Our Bayesian analysis is relevant to this program also.  Moreover, it is not limited to problems involving the pair correlation function.  For example, it can be applied to recent work proposing a polynomial parametrization of the star formation history of galaxies \citep{polynomialSFR}.  We will explore some of these ideas in forthcoming work.

\section*{Acknowledgments}
AP and RKS thank ICTP, Trieste for hospitality during the summer of 2022, when most of this work was completed.
We are grateful to F. Nikakhtar and A. Cuesta for providing the 2pcf measurements in the HADES simulations, and in BOSS and the associated QPM mock halo catalogs, respectively, that were used in this work. We thank the anonymous referee for a helpful report.

\section*{Data availability}
The numerical values of the \emph{a posteriori} mean and covariance matrix of all models explored in this work will be shared upon reasonable request to AP. 
The BOSS DR12 galaxy data and associated QPM mock catalogs are publicly available at \url{https://data.sdss.org/sas/dr12/boss/lss/}.

\bibliography{references}
 
\appendix
\section{}
\subsection{Derivation of equation~(\ref{eq:-2lnp(y|H):simple})}
\label{app:derivn}
Straightforward Gaussianology shows that $p(\yy|\Cal{H})$ is the $N$-variate Gaussian given by
\begin{align}
p(\yy|\Cal{H}) &= \int\der\aa\,p(\yy|\aa,\Cal{H})p(\aa|\Cal{H}) \notag\\
&\to \Cal{N}\left(\Cal{M}\,\aap{},C + \Cal{M}\,\Fp{}^{-1}\,\Cal{M}^{\rm T}\right)\,,
\end{align}
so that we have
\begin{align}
&-2\ln p(\yy|\Cal{H}) = \ln {\rm det}\left(C + \Cal{M}\,\Fp{}^{-1}\,\Cal{M}^{\rm T}\right) + N\ln(2\pi) \notag\\
&\ph{=p(y)}
+\left(\yy-\Cal{M}\,\aap{}\right)^{\rm T}\left(C + \Cal{M}\,\Fp{}^{-1}\,\Cal{M}^{\rm T}\right)^{-1}\left(\yy-\Cal{M}\,\aap{}\right)\,.
\label{eq:-2lnp(y|H)}
\end{align}
To simplify this, it is useful to consider the \emph{a posteriori} probability density $p(\aa|\yy,\Cal{H})$ for the parameters under hypothesis \Cal{H}, given the observed data \yy,
\be
p(\aa|\yy,\Cal{H}) = \frac{p(\yy|\aa,\Cal{H})\,p(\aa|\Cal{H})}{p(\yy|\Cal{H})}\,,
\ee
so that, using \eqn{eq:-2lnp(y|H)},
\begin{align}
&-2\ln p(\aa|\yy,\Cal{H}) \notag\\
&= \left(\yy-\Cal{M}\,\aa\right)^{\rm T} C^{-1}\left(\yy-\Cal{M}\,\aa\right) + \ln{\rm det}\,C \notag\\
&\ph{\yy} 
+ \left(\aa-\aap{}\right)^{\rm T}\Fp{}\left(\aa-\aap{}\right) - \ln{\rm det}\,\Fp{} \notag\\
&\ph{\yy\yy}
- \left(\yy-\Cal{M}\,\aap{}\right)^{\rm T} \left(C + \Cal{M}\,\Fp{}^{-1}\,\Cal{M}^{\rm T}\right)^{-1} \left(\yy-\Cal{M}\,\aap{}\right) \notag \\
&\ph{\yy\yy\yy}
 - \ln{\rm det}\left(C + \Cal{M}\,\Fp{}^{-1}\,\Cal{M}^{\rm T}\right) + M\ln(2\pi)\,.
\end{align}
The Woodbury identity gives us
\begin{align}
&\left(C + \Cal{M}\,\Fp{}^{-1}\,\Cal{M}^{\rm T}\right)^{-1} \notag\\
&= C^{-1} - C^{-1}\Cal{M}\left(\Fp{} + \Cal{M}^{\rm T}C^{-1}\Cal{M}\right)^{-1}\Cal{M}^{\rm T}C^{-1}\,,
\label{eq:Woodbury}
\end{align}
which makes it convenient to define the $M\times M$ matrix $F$ and the $M$-vector \ahat{}\ using
\begin{align}
F &\equiv \Cal{M}^{\rm T}C^{-1}\Cal{M} + \Fp{}\,, \label{eq:F-def}\\
\ahat{} &\equiv F^{-1}\left(\Cal{M}^{\rm T}C^{-1}\yy + \Fp{}\,\aap{}\right)\,. \label{eq:ahat-def}
\end{align}
Combining this with the determinant identity
\be
\ln{\rm det}\left(C + \Cal{M}\,\Fp{}^{-1}\,\Cal{M}^{\rm T}\right) = \ln{\rm det}\,C - \ln{\rm det}\,\Fp{} + \ln{\rm det}\,F\,,
\label{eq:detId}
\ee
which can be proved using the Woodbury identity along with Sylvester's theorem, straightforward algebra leads to
\be
-2\ln p(\aa|\yy,\Cal{H}) = \left(\aa - \ahat{}\right)^{\rm T}F\left(\aa - \ahat{}\right) - \ln{\rm det}\,F + M\ln(2\pi)\,,
\ee
i.e., the \emph{a posteriori} probability density of the parameters defines a Gaussian with mean \ahat{}\ and inverse covariance $F$:
\be
p(\aa|\yy,\Cal{H}) \to \Cal{N}\left(\ahat{},F^{-1}\right)\,.
\label{eq:p(a|y,H)}
\ee
The determinant identity \eqref{eq:detId} and Woodbury's identity \eqref{eq:Woodbury} can similarly be used to derive \eqn{eq:-2lnp(y|H):simple} from \eqn{eq:-2lnp(y|H)}.

\subsection{Sensitivity to range of length scales}
\label{app:scales}
In this work, we have followed the literature and chosen a `reasonable' range of values of $s$ for fitting  
$\xi(s)$ measurements \citep{LPmocks,LPnus,nsz21a}. To assess how sensitive our results are to this choice, we have varied the end-points of this range and repeated the entire Bayesian selection analysis of the BOSS CMASS data.

\begin{figure}
\centering
\includegraphics[width=0.45\textwidth]{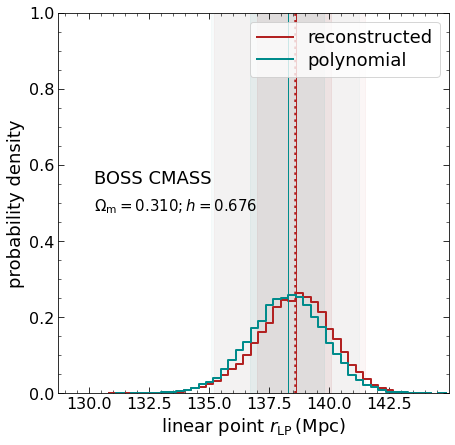}
\caption{Same as Fig.~\ref{fig:scales_boss}, but using 2pcf measurements in the range $60/0.7\leq s/(\Mpc)\leq130/0.7$ so as to match the analysis in \citet{LPboss}. In this case, the Bayesian evidence analysis selected a degree 5 Laguerre function/polynomial, which was used to infer $r_{\rm LP}$ from the corresponding reconstructed $\xi_{\rm L}$.}
\label{fig:scales_boss_broad}
\end{figure}

In general, we find that increasing the range results in higher degree Laguerre functions and polynomials being favoured by the Bayesian selection. This is sensible, since the larger range implies a larger variation of $\xi$ that must be modelled. For example, to match the analysis of \citet{LPboss}, we considered the range $60/0.7\leq s/(\Mpc)\leq130/0.7$ \citep[using the QPM cosmology values of $s$ reported by][]{cuesta+16}, leading to 35 data points. The Bayesian selection now favours degree 5 for both Laguerre as well as polynomial fitting of $\xi$ (compared to degree 3 functions selected with the narrower range used in the main text). A degree 5 polynomial was also used by \citet{LPboss} in their estimate of $r_{\rm LP}$.

Fig.~\ref{fig:scales_boss_broad} shows the result of recovering $r_{\rm LP}$ from the Laguerre reconstruction (median and $68\%$ interval $r_{\rm LP} = 138.6^{+1.5}_{-1.6}\,\Mpc$) and polynomial fit ($r_{\rm LP} = 138.3^{+1.5}_{-1.6}\,\Mpc$). These median values, which are slightly different from the ones reported in the main text, nearly coincide with the theoretical expectation of $r_{\rm LP}=138.6\,\Mpc$ in the BOSS cosmology. Moreover, the polynomial fit median value and $68\%$ interval, when scaled by $D_V$ of the BOSS cosmology to get $y_{\rm LP} = r_{\rm LP}/D_V = 0.06716^{+0.00073}_{-0.00078}$, agree extremely well with the corresponding numbers reported by \citet{LPboss} before applying their $0.5\%$ correction: $y_{\rm LP}^{\rm (Anselmi+)} = 0.06690\pm0.00073$. 
We have also explored the sensitivity of the Laguerre reconstruction to the choice of range of $s$ in larger volume samples, using the 20 HADES realisations discussed in section~\ref{sec:simrecon}. In this case, the differences in selected degree and resulting recovery of $r_{\rm LP}$ are somewhat larger. This motivates the fully Bayesian analysis alluded to in the main text, which we will report in a future publication.

% BOSS 60-130
% reconstructed
% central 68%
% r_pk : 154.65 + 1.65 - 1.79 Mpc
% r_dip: 122.56 + 1.89 - 2.01 Mpc
% r_LP : 138.61 + 1.47 - 1.62 Mpc
% central 95%
% r_pk : 154.65 + 3.18 - 3.82 Mpc
% r_dip: 122.56 + 3.77 - 4.09 Mpc
% r_LP : 138.61 + 2.86 - 3.41 Mpc

% nonlinear-poly
% central 68%
% r_LP : 138.26 + 1.50 - 1.57 Mpc
% central 95%
% r_LP : 138.26 + 2.96 - 3.20 Mpc

\label{lastpage}

\end{document}